\documentclass[journal,twoside,web]{ieeecolor}
\usepackage{tmi}
\usepackage{cite}
\usepackage{amsmath,amssymb,amsfonts}
\usepackage{algorithmic}
\usepackage{epsfig}
\usepackage{graphicx}
\usepackage{textcomp}
\usepackage{algorithm}
\usepackage{algorithmic}
\usepackage{hyperref}
\usepackage{multirow}
\usepackage{booktabs}
\usepackage{graphicx}
\usepackage{colortbl}
\usepackage{amsmath}

\def\BibTeX{{\rm B\kern-.05em{\sc i\kern-.025em b}\kern-.08em
    T\kern-.1667em\lower.7ex\hbox{E}\kern-.125emX}}
\begin{document}
\title{SliceMamba with Neural Architecture Search for Medical Image Segmentation}
\author{Chao Fan, Hongyuan Yu, Yan Huang, Liang Wang \IEEEmembership{Fellow, IEEE}, Zhenghan Yang, and Xibin Jia
\thanks{This work is partly supported by National Natural Science Foundation of China No.62171298.}
\thanks{Chao Fan and Xibin Jia (Corresponding author) are with the School of Computer Science, Beijing University of Technology, Beijing 100124, China (e-mail: chao.fancripac@gmail.com; jiaxibin@bjut.edu.cn).}
\thanks{Hongyuan Yu is with the Multimedia Department, Xiaomi Inc, Beijing 100085, China (e-mail: yuhyuan1995@gmail.com).}
\thanks{Zhenghan Yang is with the Department of Radiology, Capital Medical University Affiliated Beijing Friendship Hospital, Beijing 100050, China (e-mail: zhenghanyang@263.net).}
\thanks{Yan Huang and Liang Wang are with the New Laboratory of Pattern Recognition, Institute of Automation, Chinese Academy of Sciences, Beijing 100190, China. (e-mail: yhuang@nlpr.ia.ac.cn; wangliang@nlpr.ia.ac.cn).}
\thanks{This work has been submitted to the IEEE for possible publication. Copyright may be transferred without notice, after which this version may no longer be accessible.}
}

\maketitle

\begin{abstract}
\label{sec:abstract}
Despite the progress made in Mamba-based medical image segmentation models, existing methods utilizing unidirectional or multi-directional feature scanning mechanisms struggle to effectively capture dependencies between neighboring positions, limiting the discriminant representation learning of local features. These local features are crucial for medical image segmentation as they provide critical structural information about lesions and organs. To address this limitation, we propose SliceMamba, a simple and effective locally sensitive Mamba-based medical image segmentation model. SliceMamba includes an efficient Bidirectional Slice Scan module (BSS), which performs bidirectional feature slicing and employs varied scanning mechanisms for sliced features with distinct shapes. This design ensures that spatially adjacent features remain close in the scanning sequence, thereby improving segmentation performance. Additionally, to fit the varying sizes and shapes of lesions and organs, we further introduce an Adaptive Slice Search method to automatically determine the optimal feature slice method based on the characteristics of the target data. Extensive experiments on two skin lesion datasets (ISIC2017 and ISIC2018), two polyp segmentation (Kvasir and ClinicDB) datasets, and one multi-organ segmentation dataset (Synapse) validate the effectiveness of our method. 
\end{abstract}

\begin{IEEEkeywords}
Vision mamba, local feature, neural architecture search, medical image segmentation
\end{IEEEkeywords}

\section{INTRODUCTION}
\label{sec:introduction}
\IEEEPARstart{M}{edical} image segmentation is a pivotal process in computer-aided diagnosis (CAD), focusing on delineating regions of interest, such as lesions or organs, from medical images. Accurate segmentation methods enable clinicians to precisely identify the morphology and characteristics of pathological areas, facilitating an informed preliminary diagnosis~\cite{cheng2016computer,wang2019abdominal,wu2022cross}. This process is critical for formulating targeted treatment plans and improving patient outcomes. 

\begin{figure}
    \centering
    \includegraphics[scale=0.5]{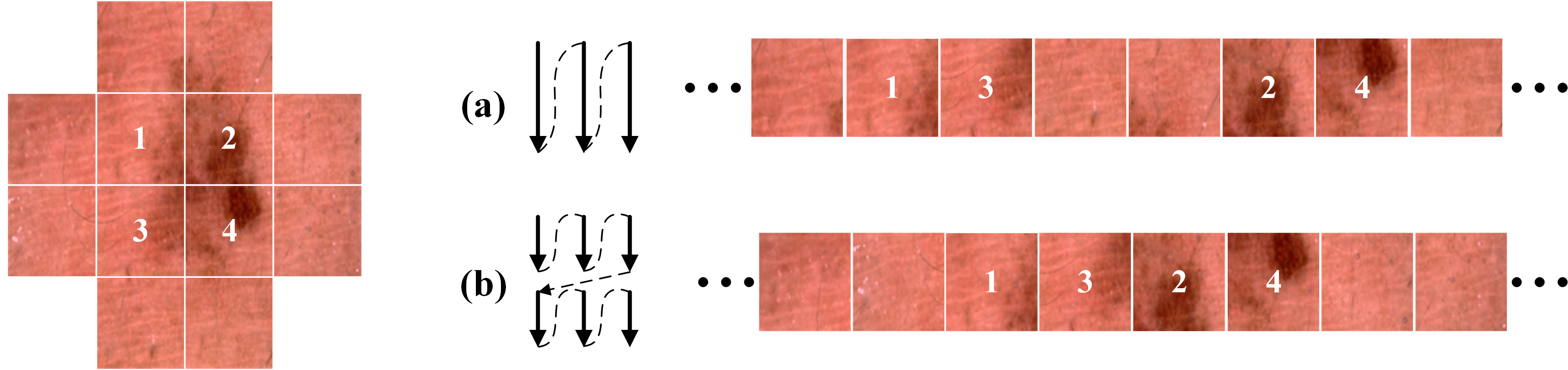}
    \caption{Results of different scanning mechanisms, illustrated with the top-to-bottom direction as an example. (a) Previous methods (VM-UNet, HC-Mamba) based on the SS2D scanning mechanism. (b) Our BSS scanning mechanism.}
    \label{fig:scan_exp}
\end{figure}

%
Recently, Mamba~\cite{gu2023mamba}, a novel State Space Model (SSM)~\cite{gu2021efficiently} in the field of natural language
processing (NLP), has emerged as a powerful long-sequence modeling approach. Unlike traditional self-attention mechanisms, Mamba enables each element (e.g., a text sequence) to interact with any previously scanned sample through a compressed hidden state, effectively reducing quadratic complexity to linear. Ma et al.~\cite{ma2024u} proposed U-Mamba, which introduced the Mamba architecture into medical image segmentation for the first time and achieved remarkable results. However, they applied the vanilla Mamba without customizing it for the inherent differences between images and sequential data, potentially limiting its effectiveness. To adapt Mamba for visual tasks, Liu et al.~\cite{liu2024vmamba} introduced VMamba. This model proposes a 2D Selective Scan (SS2D) module, tailored for spatial domain traversal, which scans input data in four directions and utilizes Mamba for modeling. Experimental results demonstrate its state-of-the-art performance across multiple perception tasks. Building upon the success of Vmamba, numerous medical image segmentation models~\cite{tang2024rotate, ruan2024vm, xu2024hc} adopt SS2D as a core component. VM-UNet~\cite{ruan2024vm} and HC-Mamba~\cite{xu2024hc} have developed pure Mamba-based medical segmentation models using the VSS block~\cite{liu2024vmamba} (with SS2D as the core component), achieving superior performance on skin lesion and Synapse datasets compared to most CNN and Transformer-based segmentation models.
However, the aforementioned Mamba-based methods fail to effectively model local features. As shown in Figure.~\ref{fig:scan_exp}(a), they extend the distance between spatially adjacent features in the scanning sequence, disrupting the local structure of the image. Consequently, this hinders further improvement in medical image segmentation performance. 
%
%

To address the aforementioned issue, we introduce SliceMamba, a locally sensitive pure Mamba-based medical image segmentation model. SliceMamba includes a simple yet effective Bidirectional Slice Scan (BSS) module, which segments features into slices along both horizontal and vertical directions. To maximize the efficiency of local feature extraction, the BSS module employs distinct scanning mechanisms that are tailored to accommodate slices of varying shapes. More specifically, horizontal features are scanned in both top-to-bottom and bottom-to-top directions, while vertical features are scanned in both left-to-right and right-to-left directions. As shown in Figure.~\ref{fig:scan_exp}(b), utilizing the BSS module ensures that spatially adjacent features in the image remain proximate in the scanning sequence. Therefore, our method can well model local features and obtain better segmentation results. 

Given the wide variety of shapes and sizes that lesions and organs can present in medical imaging, the process of manually configuring feature slicing methods can prove to be quite time-consuming and may not necessarily lead to optimal performance outcomes. To tackle this challenge, we propose an Adaptive Slice Search method to automatically select the most effective slicing strategy based on the characteristics of the target dataset. Considering both computational resources and performance, we employ the Single Path One-Shot (SPOS)~\cite{guo2020single} NAS for this task. Specifically, we define a candidate set $S$ comprising various feature slicing combinations and construct a Supernet for training. Once training is completed, we utilize an evolutionary algorithm to identify the optimal slicing method. Finally, we train and evaluate the selected method on the same dataset from scratch.

We evaluate our proposed SliceMamba on five medical image segmentation datasets with different modalities (ISIC2017~\cite{codella2018skin}, ISIC2018~\cite{codella2019skin}, Kvasir~\cite{jha2020kvasir}, ClinicDB~\cite{bernal2015wm} and Synapse~\cite{landman2015miccai}). Results show that our method can achieve leading performance, achieving $81.70\% / 89.93\% $ mIoU/DSC on the ISIC2017 dataset, $82.32\% / 90.3\%$ mIoU/DSC on the ISIC2018 dataset, $87.37\% / 93.26\%$ mIoU/DSC on the Kvasir dataset, $89.79\% / 94.61\%$ mIoU/DSC on the ClinicDB dataset, and $81.95\% / 16.04$ DSC/HD95 on the Synapse dataset.



Our contributions are summarized as follows.
\begin{itemize}
    \item We introduce SliceMamba, a pure Mamba-based model for medical image segmentation. 

    \item We introduce the Bidirectional Slice Scan (BSS) module, which incorporates an innovative feature slicing and scanning mechanism that allows Mamba to well model the local features of medical images.

    \item We propose an Adaptive Slice Search method that automatically determines the optimal feature slicing strategy according to the characteristics of the target dataset, further enhancing segmentation performance.

    \item We conduct extensive experiments across five medical image segmentation datasets with diverse modalities, and the results show that SliceMamba achieves leading performance.
\end{itemize}

\section{RELATED WORK}
\label{sec:relatedwork}
In this section, we primarily introduce the concepts related to our work, including Medical Image Segmentation and Neural Architecture Search (NAS).

\subsection{Medical Image Segmentation}
\subsubsection{CNN-based and Transformer-based Medical Image Segmentation} 
In recent years, segmentation methods based on Convolutional Neural Networks (CNN)~\cite{ronneberger2015u, zhou2019unet++, isensee2021nnu, oktay1804attention} and Transformer structures~\cite{chen2021transunet, huang2022missformer, zhang2021transfuse, he2023h2former, aghdam2023attention, hu2023devil} have gained prominence in medical image segmentation due to their exceptional performance. 
UNet~\cite{ronneberger2015u} is a groundbreaking work in medical image segmentation that introduces a symmetric encoder-decoder structure with skip connection. These connections integrate local information from the encoder with semantic features from the decoder, producing richer features that significantly improve segmentation results. Renowned for its simplicity and effectiveness, UNet has garnered considerable attention from researchers and inspired many subsequent works~\cite{zhou2019unet++, oktay1804attention, thomas2020multi, le2023rrc}. However, these CNN-based segmentation methods are constrained by their local receptive field, which hinders their ability to model long-range dependencies and results in sub-optimal segmentation outcomes.

Inspired by the success of Vision Transformers (ViTs)~\cite{dosovitskiy2020image, liu2021swin} in vision tasks, Chen et al.~\cite{chen2021transunet} introduced TransUNet, pioneering the use of transformers in medical image segmentation. TransUNet retains the overall structure of UNet, with the key difference being the replacement of ConvNets with transformers in the encoder stage to model global context. 
To further investigate the potential of transformer models in medical image segmentation, Cao et al.~\cite{cao2022swin} proposed Swin-UNet, which utilizes a pure transformer architecture as the main structure for medical image segmentation. Furthermore, He et al.~\cite{he2023h2former} proposed a hybrid CNN-Transformer architecture that combines the strengths of both Convolutional Neural Networks (CNN) and Transformers to enhance segmentation performance. While Transformers are effective at capturing long-range information, their self-attention mechanism leads to computational complexity that grows quadratically with the input size. This poses significant challenges, especially in the context of pixel-level reasoning for medical image segmentation, impacting the feasibility of transformer-based methods in practical applications.

\subsubsection{Mamba-based Medical Image Segmentation} Recent advances in State Space Models (SSMs)~\cite{gu2021efficiently}, particularly Mamba~\cite{gu2023mamba}, have demonstrated the ability to model long-range dependencies while maintaining linear complexity in terms of input size, exhibiting superior performance in a variety of vision tasks~\cite{liu2024vmamba, zhu2024vision, ruan2024vm, liu2024swin, ma2024u}. 
U-Mamba~\cite{ma2024u} introduces a novel hybrid model that combines CNN with State Space Models to capture local fine-grained features and long-range contextual information simultaneously. In this architecture, features extracted from the CNN are flattened into a 1D sequence, subsequently processed by Mamba to extract global features. Unlike natural language data, images do not have inherent causal relationships. Thus, Hao et al.~\cite {hao2024t} proposed T-Mamba, which scans the features forward and backward to improve the modeling of the image, achieving state-of-the-art results in tooth segmentation tasks.
Inspired by the success of VMamba\cite{liu2024vmamba}, numerous studies~\cite{ruan2024vm, liu2024swin, xu2024hc} on medical image segmentation have incorporated the VSS block from VMamba as a core component of their models. The key feature of the VSS block is the Cross-Scan Module (CSM), or SS2D, which scans images from four directions. This design effectively facilitates 1D scanning for 2D images, significantly improving the performance of medical image segmentation models.
However, these Mamba-based segmentation methods neglect the inherent spatial correlation in images, causing originally adjacent spatial features to become dispersed when flattened into a sequence. This dispersal hinders the modeling of local fine-grained features, ultimately degrading medical image segmentation performance. In contrast, we propose a simple yet effective method that captures local features and long-range information simultaneously without adding additional parameters.

\subsection{Neural Architecture Search}
Neural Architecture Search (NAS) represents a pivotal advancement within AutoML~\cite{he2021automl}, aimed at automatically designing optimal neural network architectures for specific tasks. NAS has proven effective for computer vision applications, including image classification~\cite{guo2020single, liu2018darts}, object detection~\cite{wang2020fcos, wu2024g}, and segmentation~\cite{xiao2022intraoperative, zhu2024saswot}. Early NAS methods~\cite{real2019regularized, zoph2018learning} involved sampling numerous architectures and then training and evaluating from scratch, which incurred unsustainable computational costs on large datasets. For instance, to obtain state-of-the-art architectures on CIFAR-10 and ImageNet, Zoph et.al~\cite{zoph2018learning} required 2000 GPU days. Recent methods~\cite{liu2018darts, pham2018efficient, bender2018understanding} adopted weight-sharing strategies to reduce the computational cost, where both the Supernet parameters and the architecture parameters are trained together. Finally, the architecture with the highest probability score is selected for training and testing from scratch on the target dataset. However, Guo et al.~\cite{guo2020single} identified severe feature coupling in these methods and proposed a Single Path One-Shot (SPOS) method using uniform sampling. By decoupling the training of the Supernet from the architecture search, SPOS achieved leading performance on various visual tasks. We optimize the ability of Mamba to extract local features of images by segmenting features bidirectionally into slices and applying various scanning mechanisms. Since lesions and organs vary in size and shape, fixed-size slices do not yield optimal segmentation results. Therefore, we propose an Adaptive Slice Search method to automatically determine the optimal feature slicing method, aiming to find suitable scanning sequences to improve image segmentation performance. scanning sequences to improve image segmentation performance.

\section{METHOD}
\label{sec:method}
In this section, we first introduce the preliminary concepts related to State Space Models (SSMs), both in continuous and discretized forms. We then detail our proposed SliceMamba, focusing on the Bidirectional Slice Scan (BSS) module and the Adaptive Slice Search method.
\subsection{Preliminaries}
State Space Models (SSMs) are a class of sequence models whose core idea is to map a one-dimensional input signal $x(t) \in \mathbb{R}$ into an output $y(t) \in \mathbb{R}$ through an intermediate latent state $h(t) \in \mathbb{R^N}$. This process can be represented by the following  linear Ordinary Differential Equation (ODE):

\begin{equation}
\label{eq:continue}
\begin{aligned}
h^{\prime}(t) & =\boldsymbol{A} h(t)+\boldsymbol{B} x(t) \\
y(t) & =\boldsymbol{C} h(t)
\end{aligned}
\end{equation}

where $\boldsymbol{A} \in \mathbb{R}^{N \times N}$, $\boldsymbol{B,C} \in \mathbb{R}^{N \times 1}$ represent the state matrix and mapping parameters, respectively.

Most modern deep learning frameworks and training algorithms are primarily designed for discrete time. Continuous-time SSMs need to be discretized to better integrate with them. The zero-order hold (ZOH) method is used by Mamba to discretize $\boldsymbol{A}$ and $\boldsymbol{B}$ in Equation \eqref{eq:continue}, resulting in the following representation:

\begin{equation}
\label{eq:discret}
\begin{aligned}
h^{\prime}(t) & =\overline{\boldsymbol{A}} h(t)+ \overline{\boldsymbol{B}} x(t) \\
y(t) & =\boldsymbol{C} h(t)
\end{aligned}
\end{equation}

where $\overline{\boldsymbol{A}}=\exp (\boldsymbol{\Delta} \boldsymbol{A})$, $\overline{\boldsymbol{B}} = (\boldsymbol{\Delta} \boldsymbol{A})^{-1}(\exp(\boldsymbol{\Delta} \boldsymbol{A})-\boldsymbol{I})\cdot \boldsymbol{\Delta}\boldsymbol{B}$, and $\boldsymbol{\Delta}$ represents a timescale parameter.

To improve computational efficiency, the iterative process in Equation \eqref{eq:discret} can be computed using global convolution, defined as follows:

\begin{equation}
\begin{aligned}
& \overline{K}=\left(\boldsymbol{C} \overline{\boldsymbol{B}}, \boldsymbol{C} \overline{\boldsymbol{A B}}, \ldots, \boldsymbol{C} \overline{\boldsymbol{A}}^{L-1} \overline{\boldsymbol{B}}\right) \\
& y=x * \overline{\boldsymbol{K}}
\end{aligned}
\end{equation}

where $*$ represents the convolution operation, and $\overline{K} \in \mathbb{R}^L$ serves as the convolutional kernel of the SSMs, $L$ denotes the length of the input sequence $x$.

\begin{figure}[h]
   \centering
   \includegraphics[scale=0.9]{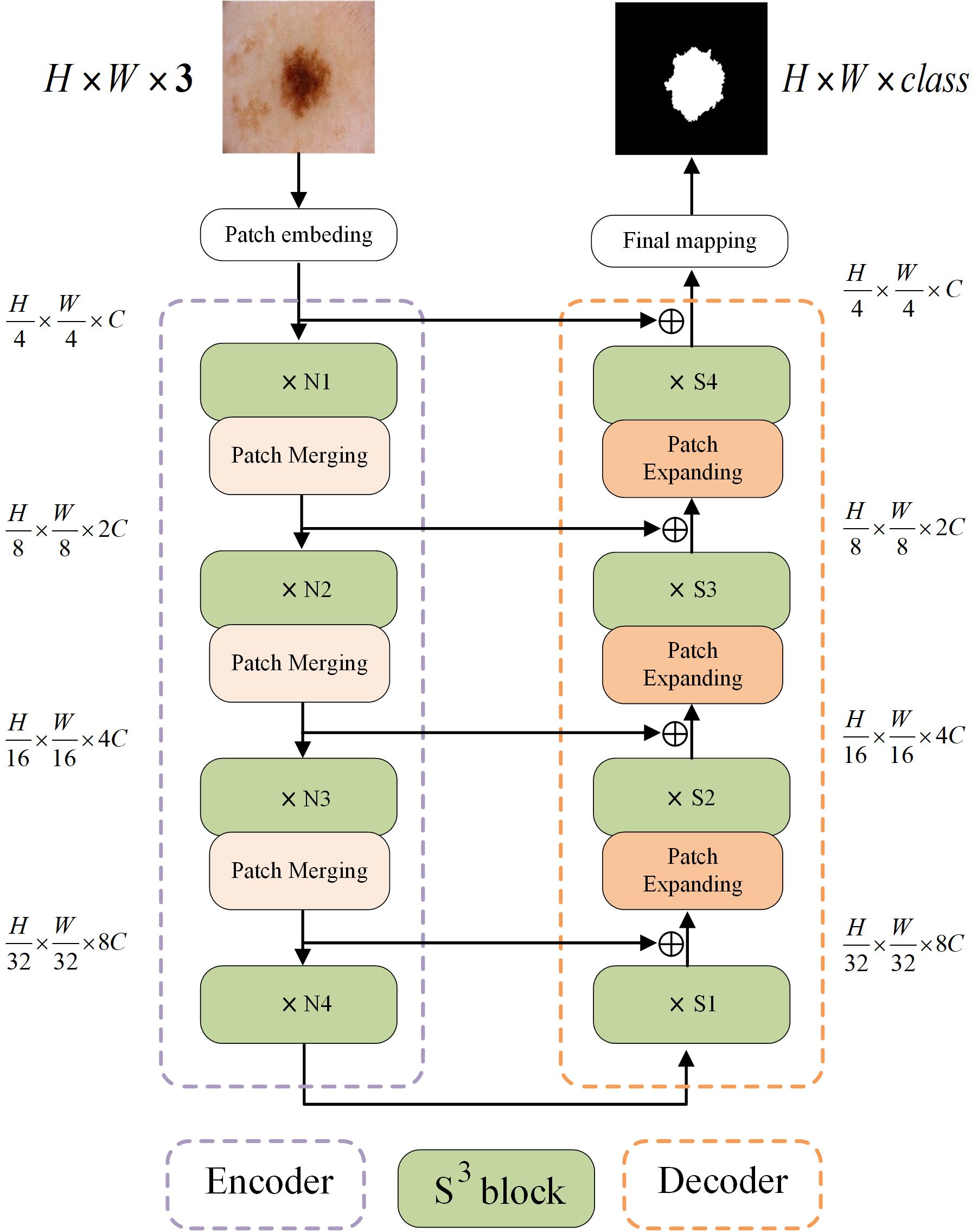}
   \caption{Overall architecture of the SliceMamba, following the classical design of UNet, comprising encoder, decoder, and skip connection operations. The details of $\mathbf{S}^3$ block are shown in Figure.\ref{fig:s3block}.}
   \label{fig:mamba-unet}
\end{figure}

\begin{figure}[h]
   \centering
   \includegraphics[scale=0.9]{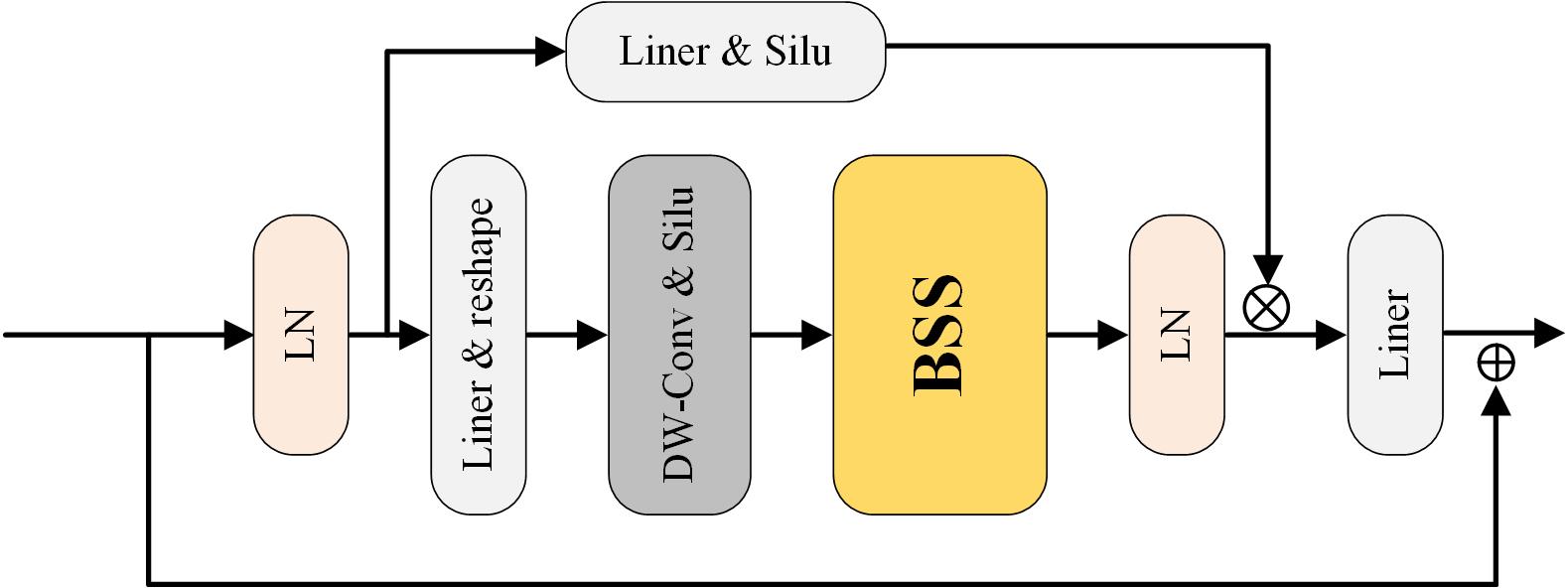}
   \caption{The $\mathbf{S}^3$ block is the main component of SliceMamba, with BSS as the core operation of the $\mathbf{S}^3$ block, which will be elaborated upon in \ref{subsec: aa-block}.}
   \label{fig:s3block}
\end{figure}

\subsection{SliceMamba}
As shown in Figure.\ref{fig:mamba-unet}, SliceMamba adopts the simple and efficient design of UNet~\cite{ronneberger2015u}. Specifically, SliceMamba includes a Patch Embedding layer, an encoder, a decoder, and a final mapping layer. In this configuration, the multi-layer features from the encoder and decoder are strengthened through a skip connection mechanism.

For example, given a skin lesion image $x \in \mathbb{R}^{H \times W \times 3}$, the patch embedding layer uses a convolution operation (with kernel size $K=4$ and stride $S=4$) to process each non-overlapping 4x4 pixel region and projects the channel dimension to $C$. Following the Layer Normalization operation, the embedded features derived from the input image are represented as $x^{\prime} \in \mathbb{R}^{\frac{H}{4} \times \frac{W}{4} \times C}$, where the $H$ and $W$ denote the height and width of the input, respectively.


Subsequently, $x^{\prime}$ is passed through the encoder for further feature extraction. The encoder consists of four stages of $\mathbf{S}^3$ blocks and Patch Merging (PM) modules. The $\mathbf{S}^3$ block maintains the shape of the input feature, while the Patch Merging module reduces the spatial dimensions of the input features by half and doubles the channel count. For the decoder, we adopt an architecture akin to that of the encoder, comprising the same stages of $\mathbf{S}^3$ blocks and Patch Expanding (PE) modules. The Patch Expanding operation in the decoder part reverses the transformation performed by the Patch Merging module in the encoder. Before each $\mathbf{S}^3$ block in the decoder, a skip connection establishes a relationship with the corresponding location in the encoder via a straightforward addition operation. Taking the middle layer $i$ of both the encoder and decoder as an example, the above operations can be formulated as follows:

\begin{equation}
\begin{aligned}
    &x^{encoder}_{i+1} = PM(S^3(x^{encoder}_{i})) \\
    &x^{decoder}_{j+1} = S^3(PE(x^{decoder}_{j}))+x^{encoder}_{L-j} \\
\end{aligned}
\end{equation}

where $x^{encoder}_{i}$ represents the features of the $i$-th layer of the encoder, and $x^{decoder}_{j}$ represents the features of the $j$-th layer of the decoder. The length of both encoder and decoder layers is $L$.

After the decoder, a Final Mapping layer is employed to resize the features to match the shape of the segmentation mask. 

\subsection{\texorpdfstring{$\mathbf{S}^3$}{S3}: Slice Selective Scan block}
\label{subsec: aa-block}
In this section, we present the structure of the $\mathbf{S}^3$ block and provide a detailed explanation of the core component, the Bidirectional Slice Scan (BSS) module, elucidating its effectiveness in modeling both local and global features.

\textbf{\textit{Notation Definition}}. $\textbf{P}^h$: horizontal slice operation, $\textbf{P}^v$: vertical slice operation, $\textbf{F}^h$: feature set of horizontal slice, $\textbf{F}^v$: feature set of vertical slice, $\textbf{Scan}^h$: top-to-bottom and bottom-to-top scanning operation, $\textbf{Scan}^v$: left-to-right and right-to-left scanning operation, $\textbf{S6}$: Selective SSM~\cite{gu2023mamba}, $\textbf{Restore}$: scan merging and shape restore operation. 

The $\mathbf{S}^3$ module is a core component of our SliceMamba, designed to capture both local and global features across different stages of feature maps. Different from previous methods~\cite{yue2024medmamba} that relied on ConvNets to assist Mamba in local feature extraction and long-range dependency modeling, our approach relies entirely on the Mamba structure to achieve both types of feature modeling in a simple yet effective manner.

\begin{figure*}
    \centering
    \includegraphics[scale=0.9]{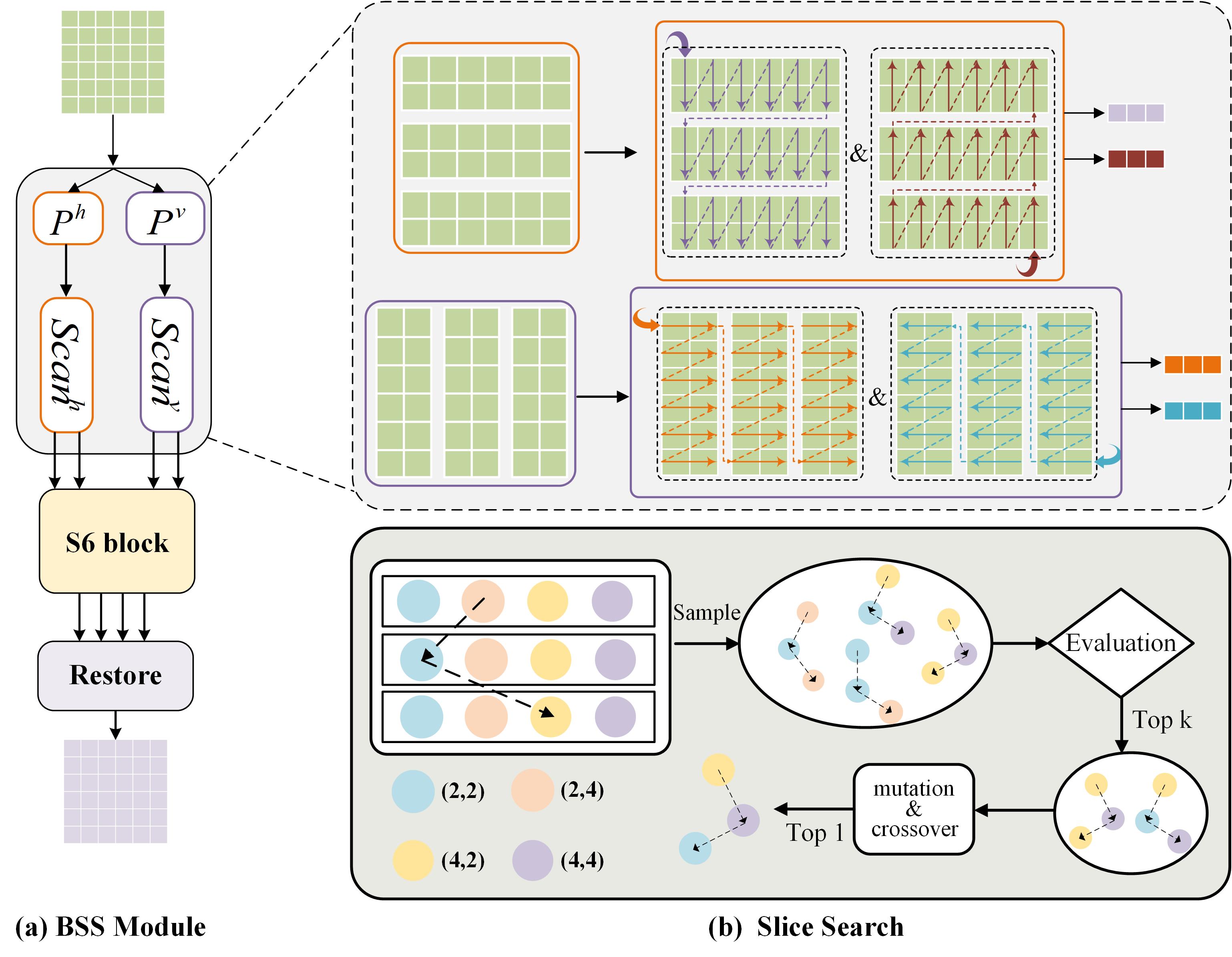}
    \caption{(a) Structure of the BSS module, which includes feature slicing and scanning operations, the S6 block, and the feature restoration process. (b) Illustration of the search process for feature slicing methods.}
    \label{fig:lgss2d}
\end{figure*}

As depicted in Figure \ref{fig:s3block}, the input features first pass through an initial linear embedding layer before being split into two paths. In one path, the features undergo a 3x3 depth-wise convolution followed by a Silu activation and then processed by the BSS module. Subsequently, the features from this path are normalized with Layer Normalization and then element-wise multiplied with the features from the other path. Finally, the product of these two paths is passed through a linear embedding layer and then connected residually to the $\mathbf{S}^3$ block input features, resulting in the final output features from the $\mathbf{S}^3$ block.

\subsection{BSS: Bidirectional Slice Scan}
\label{sec:BSS}

The BSS module comprises four components: a bidirectional feature slicing operation, a scan expanding operation, an S6 block, and a restore operation. As shown in Figure.\ref{fig:lgss2d}(a), we consider an input feature $F\in \mathbb{R}^{H\times W \times C}$, where $H$, $W$, and $C$ represent the height, width, and channel count, respectively. The feature $F$ first undergoes bidirectional feature slicing, which involves slicing in both horizontal and vertical directions. Specifically, $F$ is segmented horizontally into $\frac{H}{m}$ horizontal slices $F^h = \left\{f^h_i \in \mathbb{R}^{m \times W \times C} \mid i=1,2, \cdots \frac{H}{m}\right\}$, and vertically into $\frac{W}{n}$ vertical slices $F^v = \left\{f^v_j \in \mathbb{R}^{H \times n \times C} \mid j=1,2, \cdots \frac{W}{n}\right\}$. Without loss of generality, we assume that $\frac{H}{m}$ and $\frac{W}{n}$ are both integers. 
Subsequently, we apply different scanning mechanisms to slices of varying shapes. Slices with a shape of $m \times W \times C$ are scanned in both top-to-bottom and bottom-to-top directions, while slices with a shape of $H \times n \times C$ are scanned in both left-to-right and right-to-left directions.
The four resulting sequences from these scans are sent to the S6 block for modeling using Mamba. The output sequences from the S6 block then undergo a restore operation, where features from corresponding positions are combined using element-wise addition and restored to the same shape as $F\in \mathbb{R}^{H\times W \times C}$. 
%

We will illustrate why the BSS module provides a simple yet effective way to extract local features and model long-range dependencies. 

\textbf{Regarding long-range modeling}. Although we segment the features into multiple slices, their continuity remains intact during the scanning phase. For convenience, we take the example of segmenting the feature into $\frac{H}{m}$ horizontal slices $F^h$ using top-to-bottom scanning. As shown in Figure.~\ref{fig:lgss2d}(a), each horizontal slice feature $f^h_i$ scanned from top-to-bottom, resulting in $\frac{H}{m}$ sequences $\left\{s^h_i \in \mathbb{R}^{(m \times W) \times C} \mid i=1,2, \cdots \frac{H}{m}\right\}$. These sequences are then concatenated along the last dimension to form a scanning sequence $S^h \in \mathbb{R}^{(m\times W \times \frac{H}{m}) \times C}$ for the entire feature, enabling global modeling. 
It is noteworthy that when $m=H$, the resulting sequences are identical to those produced by other methods utilizing the SS2D scanning mechanism. Therefore, the SS2D scanning mechanism is considered a special case of our proposed method.
\begin{figure*}
    \centering
    \includegraphics[width=0.95\textwidth]{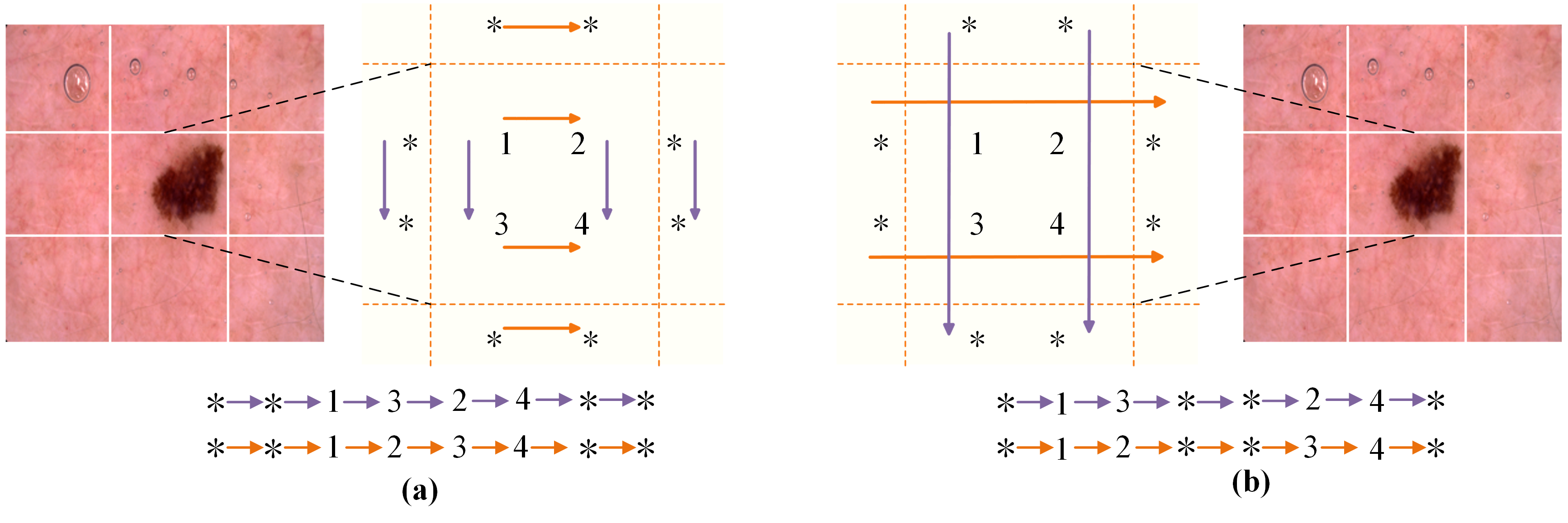}
    \caption{We compare our scanning method with previous works,  (a) The scanning results of our SliceMamba, (b) The scanning results of previous works, such as VM-UNet~\cite{ruan2024vm} and TM-UNet~\cite{tang2024rotate}}.
    \label{fig:local-exp}
\end{figure*}

\textbf{For local feature modeling}. For clarity, we only use the example of top-to-bottom and left-to-right scanning mechanisms. As shown in Figure.~\ref{fig:local-exp}, let $\left[\begin{array}{ll}1 & 2 \\ 3 & 4\end{array}\right]$ represent the overlap area of two types of slice feature $f^h_i$ and $f^v_j$, and let $*$ represent the surrounding pixels. In Figure ~\ref{fig:local-exp}(a), scanning is conducted top-to-bottom on the horizontal slice feature and left-to-right on the vertical slice feature, resulting in two corresponding scanning sequences. Due to this novel design, the original spatially adjacent features in space maintain proximity in the scanning sequence, facilitating local feature modeling. In contrast, other Mamba-based methods are shown in Figure.~\ref{fig:local-exp}(b), increasing the distance between adjacent features in the scanning sequence, and hindering local feature modeling. 

\subsection{Adaptive Slice Search}
In Section~\ref{sec:BSS}, the BSS module performs both horizontal and vertical slicing on the input features and applies corresponding scanning mechanisms. This ensures that features in adjacent spatial positions within overlapping areas remain close in scanning sequences. Since different lesions or organs have varying sizes and shapes, it is important to adapt the shape of the overlapping regions to the target to enhance the representation learning of local features without introducing excessive background. To address this, we introduce an Adaptive Slice Search method to automatically select the optimal feature slicing method, enhancing performance without adding any extra network parameters.
\subsubsection{Search Space}
Considering the general shapes and sizes of lesions and organs, we introduce a set \( S = \{ (2, 2), (2, 4), (4, 2), (4, 4) \} \), comprising 4 candidate feature slicing combinations. Taking \( (2, 4) \) for example, the height of each horizontal slice feature is 2 ($f^h_i \in \mathbb{R}^{2 \times W \times C}$), and the width of each vertical slice feature is 4 ($f^v_j \in \mathbb{R}^{H \times 4 \times C}$). To achieve more diverse slice combinations, each BSS module in an $\mathbf{S}^3$ block selects one slice method from the set. This approach generates a substantial search space of \( 4^k \), where \( k \) is the total number of blocks.

\subsubsection{Search method}
Currently, two widely used NAS methods are Differentiable
ARchiTecture Search (DARTS)~\cite{liu2018darts} and Single Path One-Shot (SPOS)~\cite{guo2020single}, both significantly reduce GPU computation time and enhance efficiency. In our work, we adopt the SPOS method because DARTS tends to exceed GPU memory limits and suffers from severe coupling between the architecture parameters and Supernet weights. SPOS decouples the Supernet training and architecture search into two distinct steps, effectively mitigating weight coupling issues and enabling better control over GPU memory usage.

The process of Supernet training and feature slice combination search in SPOS is conducted in two steps. First, the optimization of Supernet weights can be formulated as
\begin{equation}
W_{\mathcal{S}}=\underset{W}{\operatorname{argmin}} \mathcal{L}_{\text {train }}(\mathcal{N}(\mathcal{S}, W))
\end{equation}

Where $\mathcal{N}(\mathcal{S}, W)$ represents the Supernet, $\mathcal{S}$ denotes the Supernet architecture that includes all candidate slicing method combinations, and $\mathcal{W}$ denotes the weights. The Supernet is trained only once, and all candidates inherit their weights directly from $\mathcal{W}$ for search.

Second, the optimal feature slice combination search process can be formulated as
\begin{equation}
c^*=\underset{c \in \mathcal{S}}{\operatorname{argmax}} \operatorname{DSC}_{\text {val }}\left(\mathcal{N}\left(c, W_{\mathcal{S}}(c)\right)\right) .
\end{equation}

Where $\mathcal{N}(c, W_{\mathcal{S}}(c))$ represents sampling a slice combination $c$ from the candidate set $S$, inheriting corresponding parameters from $W_\mathcal{S}$ as $W_{\mathcal{S}}(c)$, and evaluating it on the validation set. To find the combination with optimal performance, we utilize an evolutionary algorithm for the search process, the overall search process is shown in Figure.~\ref{fig:lgss2d}(b). Finally, we train the searched combination from scratch on the training dataset and evaluate it on the test dataset.

\section{EXPERIMENTS}
\label{sec:experiments}
\subsection{Datasets}
To validate the effectiveness of the proposed SliceMamba, we conduct experiments on two skin lesion segmentation datasets, two polyp segmentation datasets, and one Multi-organ segmentation dataset.

\subsubsection{Skin Lesion datasets} \textbf{ISIC2017}~\cite{codella2018skin} and \textbf{ISIC2018}~\cite{codella2019skin} are two publicly available skin lesion segmentation datasets published by the International Skin Imaging Collaboration (ISIC). The ISIC2017 dataset contains 2150 images, while the ISIC2018 dataset includes 2694 images, each with its corresponding lesion segmentation mask. Following VM-UNet~\cite{ruan2024vm}, we split the datasets into training and testing sets in a 7:3 ratio. For ISIC2017, 1500 images are allocated for training and 650 for testing. For ISIC2018, 1886 images are allocated for training, and 808 images are reserved for testing.

\subsubsection{Polyp datasets}
\textbf{Kvasir}~\cite{jha2020kvasir} and \textbf{ClinicDB}~\cite{bernal2015wm} are two open-source gastrointestinal polyp datasets. Kvasir contains 1000 high-resolution color polyp images, and ClinicDB has 612, both with manually annotated segmentation masks by clinicians. We randomly selected 900 images from Kvasir and 550 from ClinicDB for training, following the setup of previous work~\cite{zou2022graph}, resulting in a total of 1450 training images. The remaining images from each dataset are used for testing.

\subsubsection{Multi-organ segmentation dataset} 
The \textbf{Synapse}~\cite{landman2015miccai} is a publicly available multi-organ segmentation dataset from the
MICCAI 2015 Multi-Atlas Abdomen Labeling Challenge. It contains 30 abdominal CT scans with 3779 axial abdominal clinical CT images, including 8 types of abdominal organs (aorta, gallbladder, left kidney, right kidney, liver, pancreas, spleen, and stomach). Following the setup of previous studies~\cite{ruan2024vm, zou2022graph}, we use 18 cases for training and the remaining 12 cases for testing.

\subsection{Implementation Details}
We conducted training and testing of our proposed model on a system running Ubuntu 22.04, with Python version 3.8.19, CUDA version 11.8, and an RTX4090 GPU. Before feeding into the model, images were resized to 256x256 and employed data augmentation including rotation, vertical flip, and horizontal flip to prevent overfitting.
\subsubsection{Supernet training and slice combination search}
\begin{figure}
    \centering
    \includegraphics[scale=0.9]{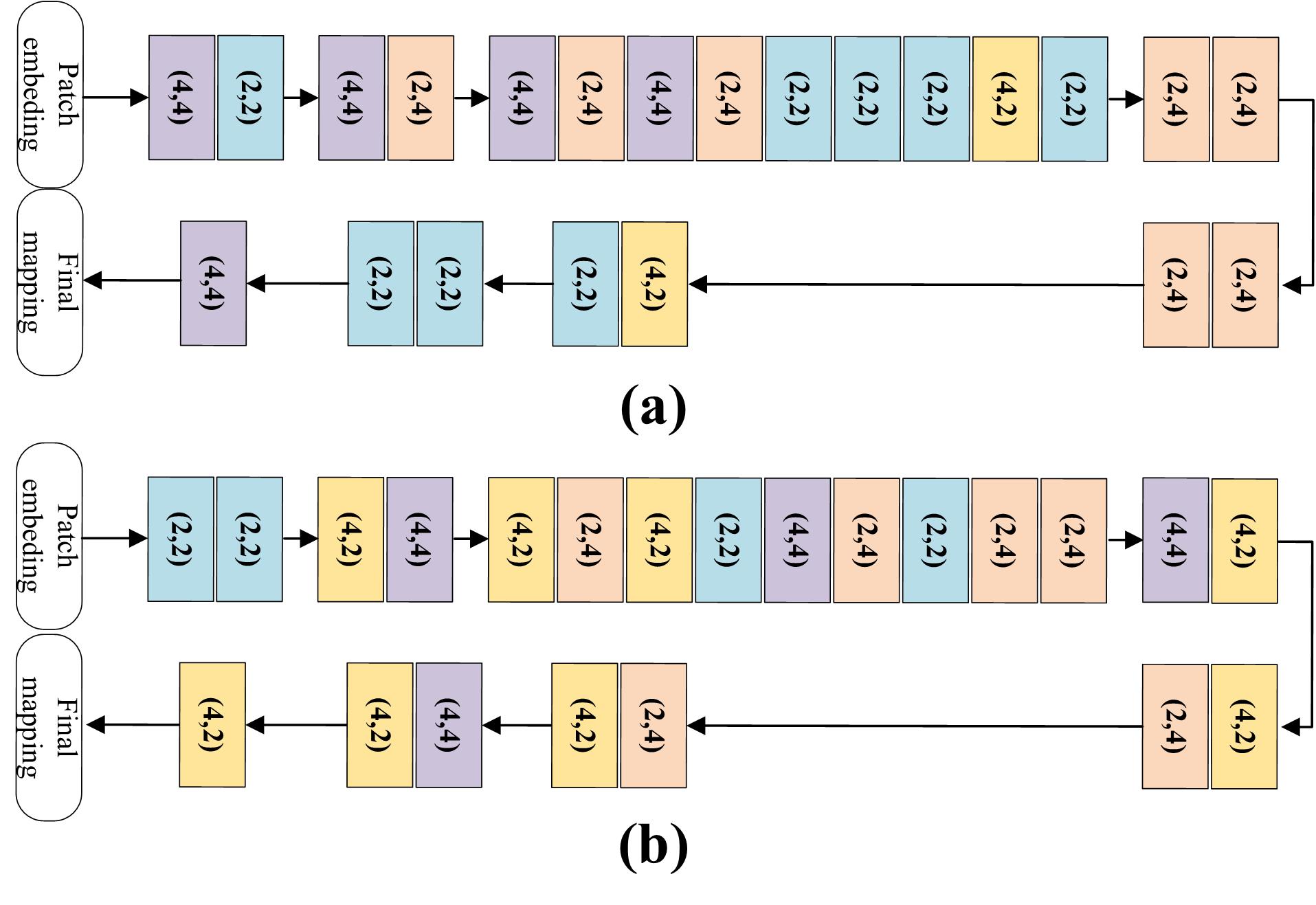}
    \caption{The visualization of the slice combinations of our model omits skip connections for clarity. Since skin lesions and polyps have similar sizes and shapes, they share the slice combination shown in (a). For organ segmentation, we searched for the combination as illustrated in (b).}
    \label{fig:search_archi}
\end{figure}

We randomly select $80\%$ of the training data for Supernet training, and the remaining $20\%$ for slice method search. The number of $\mathbf{S}^3$ blocks in each layer of the encoder and decoder are set to [2,2,9,2] and [2,2,2,1] respectively, and the feature slice candidate set in each block is \( S = \{ (2, 2), (2, 4), (4, 2), (4, 4) \} \). For training details, we employ the AdamW~\cite{loshchilov2017decoupled} with an initial learning rate of 1e-3, coupled with the CosineAnnealingLR~\cite{loshchilov2016sgdr} scheduler, which iterates a maximum of 50 times and reaches a minimum learning rate of 1e-5. We maintain a uniform batch size of 32 for all datasets and train for 300 epochs. The BceDice loss (Binary Cross-Entropy and Dice loss) is utilized for skin lesion and Polyp datasets, while the CeDice loss (Cross-Entropy and Dice loss) is adopted for the Synapse
dataset. Once the Supernet training is complete, we freeze its parameters and use an evolutionary algorithm to search for feature segmentation combinations that exhibit leading performance on the search dataset. As shown in Figure.~\ref{fig:lgss2d} (b), we randomly sample 50 combinations for validation, then select the top 10 based on evaluation metrics for crossover and mutation. This process is repeated 20 times, and the combination with the best performance is saved. We visualize the searched slice combinations of our model in Figure.~\ref{fig:search_archi}.
\subsubsection{Training from scratch}
We first pre-train the encoder part of the architecture obtained through the Adaptive Slice Search method on ImageNet-1K~\cite{deng2009imagenet}. After pre-training, we initialize both the encoder and decoder with the obtained parameters and train from scratch on the entire training dataset. The training details are identical to those used in the Supernet training.

\begin{table}[h]
	\begin{center}
		\caption{Quantitative segmentation results on the ISIC2017 skin lesion dataset.}
            \vspace{5pt}
		\scalebox{1}{
			\begin{tabular}{l rrrrc rrrrc}
				\hline 
				\multicolumn{1}{c}{\multirow{2}{*}{ Methods}} & \multicolumn{5}{c}{ISIC2017}  \\
                    \cmidrule(lr){2-6}
			& \multicolumn{1}{c}{mIoU$\uparrow$} & \multicolumn{1}{c}{DSC$\uparrow$} & \multicolumn{1}{c}{Acc$\uparrow$} & \multicolumn{1}{c}{Spe$\uparrow$} & \multicolumn{1}{c}{Sen$\uparrow$} \\
				\hline 
                    UNet~\cite{ronneberger2015u}
                    & 76.98 & 86.99 & 95.65 & 97.43   & 86.82  \\
            
                    UNet++~\cite{zhou2019unet++}
                    & 75.44 & 86.00 & 95.35 & 97.34   & 85.40 \\
                    
                    Att-UNet~\cite{oktay1804attention}
                    & 76.17 & 86.47 & 95.49 & 97.40   & 86.02 \\

                    TransUNet~\cite{chen2021transunet}
                    & 78.79 & 88.13 & 96.12 & 98.14   & 86.05 \\
                   
                    TransFuse~\cite{zhang2021transfuse}
                    & 79.21 & 84.40 & 96.17 & 97.98   & 87.14 \\

                    Att-Swin UNet~\cite{aghdam2023attention}
                    & 76.32 & 86.28 & 94.15 & 95.66   & 89.44\\

                    $\mathbf{C}^2 \text {SGD}$~\cite{hu2023devil}
                    & 79.18 & 89.09 & 95.88 & 97.65   & 88.59  \\

                    VM-UNet~\cite{ruan2024vm}
                    & 80.23 & 89.03 & 96.29 & 97.58   & 89.90  \\

                    TM-UNet~\cite{tang2024rotate}
                    & 80.51& 89.20 & 96.46 & 98.28   & 87.37  \\

                    HC-Mamba~\cite{xu2024hc}
                    &79.27 &  88.18 & 95.17 & 97.47 &86.99 \\
                    \rowcolor[gray]{0.9}
                    SliceMamba
                    & \textbf{81.70} & \textbf{89.93} & 96.75 & 98.3   & 88.81 \\
           
				\hline 
		\end{tabular}
          }
		\label{skin_lesion_table_17}
	\end{center}
\end{table}

\begin{table}[h]
	\begin{center}
		\caption{Quantitative segmentation results on the ISIC2018 skin lesion datasets}
            \vspace{5pt}
		\scalebox{1}{
			\begin{tabular}{l rrrrc rrrrc}
				\hline 
				\multicolumn{1}{c}{\multirow{2}{*}{\centering Methods}} & \multicolumn{5}{c}{ISIC2018}  \\
                    \cmidrule(lr){2-6}
			& \multicolumn{1}{c}{mIoU$\uparrow$} & \multicolumn{1}{c}{DSC$\uparrow$} & \multicolumn{1}{c}{Acc$\uparrow$} & \multicolumn{1}{c}{Spe$\uparrow$} & \multicolumn{1}{c}{Sen$\uparrow$}  \\
				\hline 
                    UNet~\cite{ronneberger2015u}
                    & 77.86 & 87.55 & 94.05 & 96.69 & 85.86 \\
            
                    UNet++~\cite{zhou2019unet++}
                    &78.31 & 87.83 & 94.02  & 95.75 & 88.65 \\
                    
                    Att-UNet~\cite{oktay1804attention}
                    &77.23 & 87.16 & 93.93  & 96.93 & 84.61 \\

                    TransUNet~\cite{chen2021transunet}
                    & 81.09 & 89.56 & 94.99 & 97.02   & 88.14 \\
                   
                    TransFuse~\cite{zhang2021transfuse}
                    & 80.63 & 89.27 & 94.66   & 95.74 & 91.28 \\

                    Att-Swin UNet~\cite{aghdam2023attention}
                    & 78.32 & 87.99 & 93.91   & 95.37 & 89.33 \\

                    $\mathbf{C}^2 \text {SGD}$~\cite{hu2023devil}
                    & 80.48 & 89.41 & 94.11   & 96.58 & 87.93 \\

                    VM-UNet~\cite{ruan2024vm}
                    &81.35 & 89.71 & 94.91   & 96.13 & 91.12 \\

                    TM-UNet~\cite{tang2024rotate}
                    & 81.55 & 89.84 & 95.08   & 96.68 & 89.98 \\

                    HC-Mamba~\cite{xu2024hc}
                    & 80.72 & 89.26 & 94.84   & 97.08 & 88.90 \\
                    \rowcolor[gray]{0.9}
                    SliceMamba
                    & \textbf{82.32} & \textbf{90.30} & 95.29   & 97.14 & 89.58 \\
 
                    
				\hline 
		\end{tabular}
          }
		\label{skin_lesion_table_18}
	\end{center}
\end{table}

\subsection{Evaluation Results}
To demonstrate the effectiveness of our proposed SliceMamba, we compare it with several state-of-the-art medical image segmentation methods, including CNN-based, Transformer-based, and Mamba-based segmentation models. 

\subsubsection{Results of Skin Lesion segmentation}
For the ISIC2017 and ISIC2018 skin lesion datasets, we employ five evaluation metrics to measure the segmentation performance of the models, including mean Intersection over Union(mIoU), dice similarity coefficient(DSC), Accuracy(Acc), specificity(Spe), and sensitivity(Sen). Comparison results with other state-of-the-art methods are shown in Table.~\ref{skin_lesion_table_17} and Table.~\ref{skin_lesion_table_18}, and SliceMamba achieves the best results in mIou and DSC metrics. Specifically, SliceMamba outperforms the state-of-the-art CNN-based method Att-UNet by $5.53\%$ and $5.09\%$ in mIoU on the ISIC2017 and ISIC2018 datasets, respectively. Additionally, it surpasses by $3.46\%$ and $3.14\%$ in terms of the DSC metric. Compared to those Transformer-based models, such as $\mathbf{C}^2$SGD, our approach also exhibits significant advantages in mIoU and DSC metrics, with improvements of $2.52\% / 0.84\%$ mIoU/DSC on ISIC2017 and $1.84\% / 0.89\%$ mIoU/DSC on ISIC2018, respectively. Furthermore, compared to the representative Mamba-based methods VM-UNet, TM-UNet, and HC-Mamba, our approach also achieves leading performance.

\begin{table}[t]
	\begin{center}
		\caption{Quantitative segmentation results on Kvasir dataset.}
            \vspace{5pt}
		\scalebox{1}{
			\begin{tabular}{l rrrrc rrrrc}
				\hline 
				\multicolumn{1}{c}{\multirow{2}{*}{\centering Methods}} & \multicolumn{5}{c}{Kvasir}  \\
				\cmidrule(lr){2-6}
			& \multicolumn{1}{c}{mIoU$\uparrow$} & \multicolumn{1}{c}{DSC$\uparrow$} & \multicolumn{1}{c}{Acc$\uparrow$} & \multicolumn{1}{c}{Spe$\uparrow$} & \multicolumn{1}{c}{Sen$\uparrow$} \\
				\hline 
                    UNet~\cite{ronneberger2015u}
                    & 72.46 & 84.03 & 95.32 & 97.65   & 82.12  \\
            
                    UNet++~\cite{zhou2019unet++}
                    & 76.17 & 86.48 & 96.11 & 98.44   & 82.89 \\
                    
                    Att-UNet~\cite{oktay1804attention}
                    & 76.05 & 86.39 & 96.17 & 98.83   & 81.09 \\

                    Polyp-PVT~\cite{dong2021polyp}
                    & 86.40 & 91.70 & - & -   & - \\

                    TransUNet~\cite{chen2021transunet}
                    & 79.59 & 88.63 & 96.73 & 98.84   & 84.81 \\

                    TransFuse~\cite{zhang2021transfuse}
                    & 68.82 & 81.53 & 94.81 & 98.07   & 76.34 \\

                    Att-Swin UNet~\cite{aghdam2023attention}
                    & 75.63 & 86.16 & 94.61 & 97.81   & 83.16 \\

                    $\mathbf{C}^2 \text {SGD}$~\cite{hu2023devil}
                    & 77.80 & 87.32 & 96.28 & 98.33   & 84.57 \\

                    VM-UNet~\cite{ruan2024vm}
                    & 78.66 & 88.05 & 96.61& 98.96   & 83.27 \\

                    TM-UNet~\cite{tang2024rotate}
                    & 77.83 & 87.53 & 96.37 & 98.37   & 85.02 \\
                    \rowcolor[gray]{0.9}
                    SliceMamba
                    & \textbf{87.37} & \textbf{93.26} & 98.01 & 99.10  & 91.80 \\
                    
				\hline 
		\end{tabular}
          }
		\label{polyp_results_kva}
	\end{center}
\end{table}

\begin{table}[t]
	\begin{center}
		\caption{Quantitative segmentation results on ClinicDB dataset.}
            \vspace{5pt}
		\scalebox{1}{
			\begin{tabular}{l rrrrc rrrrc}
				\hline 
				\multicolumn{1}{c}{\multirow{2}{*}{\centering Methods}} & \multicolumn{5}{c}{ClinicDB}  \\
				\cmidrule(lr){2-6}
			& \multicolumn{1}{c}{mIoU$\uparrow$} & \multicolumn{1}{c}{DSC$\uparrow$} & \multicolumn{1}{c}{Acc$\uparrow$} & \multicolumn{1}{c}{Spe$\uparrow$} & \multicolumn{1}{c}{Sen$\uparrow$} \\
				\hline 
                    UNet~\cite{ronneberger2015u}
                    & 78.60 & 88.02 & 98.13 & 99.19   & 85.92  \\
            
                    UNet++~\cite{zhou2019unet++}
                    & 83.24 & 90.86 & 98.59 & 99.53   & 87.66 \\
                    
                    Att-UNet~\cite{oktay1804attention}
                    & 79.46 & 88.55& 98.26 & 99.50   & 84.00 \\

                    Polyp-PVT~\cite{dong2021polyp}
                    & 88.90 & 93.70 & - & -   & - \\

                    TransUNet~\cite{chen2021transunet}
                    & 81.45 & 89.77 & 98.44 & 99.57   & 85.39 \\

                    TransFuse~\cite{zhang2021transfuse}
                    & 71.59 & 83.44 & 97.51 & 99.14   & 78.69 \\

                    Att-Swin UNet~\cite{aghdam2023attention}
                    & 79.66 & 88.68 & 98.22 & 99.18   & 87.22 \\

                    $\mathbf{C}^2 \text {SGD}$~\cite{hu2023devil}
                    & 83.30 & 91.01 & 97.71 & 99.17   & 91.06 \\

                    VM-UNet~\cite{ruan2024vm}
                    & 76.53 & 86.71 & 97.92& 99.05   & 84.89 \\

                    TM-UNet~\cite{tang2024rotate}
                    & 83.82 & 91.20 & 98.59 & 99.25   & 91.10 \\
                    \rowcolor[gray]{0.9}
                    SliceMamba
                    & \textbf{89.78} & \textbf{94.61} & 99.12 & 99.41   & 95.81 \\
                    
				\hline 
		\end{tabular}
          }
		\label{polyp_results_cli}
	\end{center}
\end{table}

\subsubsection{Results of Polyp segmentation}
Comparison results with state-of-the-art methods on two polyp segmentation datasets are shown in Table.~\ref{polyp_results_kva} and Table.~\ref{polyp_results_cli}, which demonstrate that our SliceMamba outperforms other methods by a large margin. Compared to the representative work Polyp-PVT~\cite{dong2021polyp} in the field of polyp segmentation, SliceMamba achieves a $0.97\%$ and $1.56\%$ improvement in mIoU on the Kvasir and ClinicDB datasets, respectively, and a $0.88\%$ and $0.91\%$ improvement in DSC on these datasets. In comparison to Mamba-based, TM-UNet, our method exhibits more notable achievements across these two datasets, with mIoU improvements of $9.54\%$ and $5.96\%$, and DSC improvements of $5.73\%$ and $3.41\%$, respectively. In contrast with skin lesion datasets, polyp datasets exhibit similarities between lesions and background, highlighting the importance of models in efficiently extracting local features. This underscores the reason why our method achieves significant improvements over other Mamba-based segmentation models on polyp datasets.

\subsubsection{Results of Synapse segmentation}
To evaluate the performance of different segmentation methods on the Synapse dataset, we use DSC and HD95 ($95\%$ Hausdorff Distance) as evaluation metrics. HD95 measures the boundary matching between the predicted mask and the ground truth, with lower values indicating better edge detail extraction by the model. As shown in Table.~\ref{Synapse_dataset}, our proposed SliceMamba outperforms the first Mamba-based model, VM-UNet, by $0.87\%$ in DSC and 3.17mm in HD95. From these results, it can be seen that although our method does not show significant improvement over VM-UNet in the DSC metric, the improvement in the HD95 metric is more pronounced, indicating that our method is effective in extracting edge detail features.

\begin{table*}[t]
	\begin{center}
		\caption{Quantitative segmentation results on the Synapse dataset.}
            \vspace{5pt}
		\scalebox{1}{
			\begin{tabular}{l cc cccccccc}
				\hline 
                \multicolumn{1}{c}{\multirow{2}{*}{\centering Methods}} & \multicolumn{2}{c}{Average performance} & \multicolumn{8}{c}{Dice of Each Class} \\
                \cmidrule(lr){2-3}\cmidrule(lr){4-11}
			& \multicolumn{1}{c}{DSC$\uparrow$} & \multicolumn{1}{c}{HD95$\downarrow$} & \multicolumn{1}{c}{Aorta} & \multicolumn{1}{c}{Gallbladder} & \multicolumn{1}{c}{Kidney(L)} & \multicolumn{1}{c}{Kidney(R)} & \multicolumn{1}{c}{Liver} & \multicolumn{1}{c}{Pancreas} & \multicolumn{1}{c}{Spleen} & \multicolumn{1}{c}{Stomach}  \\
				\hline 
                    V-Net~\cite{milletari2016v}
                    & 68.81 & - & 75.34 & 51.87   & 77.10 & 80.75 & 87.84 & 40.05 & 80.56 & 56.98 \\

                    UNet ~\cite{ronneberger2015u}
                     & 76.85 & 39.70 & 89.07 & 69.72   & 77.77 & 68.60 & 93.43 & 53.98 & 86.67 & 75.58 \\
                    
                    Att-UNet~\cite{oktay1804attention}
                    & 77.77 & 36.02 & 89.55 & 68.88   & 77.98 & 71.11 & 93.57 & 58.04  & 87.30 & 75.75 \\

                    Graph-Flow~\cite{zou2022graph}
                    & 78.29 & 26.99 & 85.11 & 62.74   & 84.62 & 81.40 & 94.51 & 54.49  & 89.82 & 73.75 \\

                    TransUNet~\cite{chen2021transunet}
                   & 77.48 & 31.69 & 87.23 & 63.13   & 81.87 & 77.02 & 94.08 & 55.86  & 85.08 & 75.62 \\
                   
                    Swin-UNet~\cite{cao2022swin}
                    & 79.13 & 21.55 & 85.47 & 66.53   & 83.28 & 79.61 & 94.29 & 56.58   & 90.66 & 76.60 \\

                    TransDeepLab~\cite{azad2022transdeeplab}
                    & 80.16 & 21.25 & 86.04 & 69.16   & 84.08 & 79.88 & 93.53 & 61.19   & 89.00 & 78.40 \\

                    MT-UNet~\cite{wang2022mixed}
                    & 78.59 & 26.59 & 87.92 & 64.99   & 81.47 & 77.29 & 93.06 & 59.46   & 87.75 & 76.81 \\

                    MEW-UNet~\cite{ruan2022mew}
                    & 78.92 & 16.44 & 86.68 & 65.32   & 82.87 & 80.02 & 93.63 & 58.36   & 90.19 & 74.26 \\

                    VM-UNet~\cite{ruan2024vm}
                    & 81.08 & 19.21 & 86.40 & 69.41   & 86.16 & 82.76 & 94.17 & 58.80   & 89.51 & 81.40 \\

                    HC-Mamba~\cite{xu2024hc}
                    & 81.58 & 26.34 & 90.93 & 69.65   & 85.57 & 79.27 & 97.38 & 54.08   & 93.49 & 80.14 \\
                    \rowcolor[gray]{0.9}
                    SliceMamba
                    & \textbf{81.95} & \textbf{16.04} & 87.78 & 68.77   &88.30 &
                    84.26 & 95.25 & 64.49  & 86.91 & 79.82 \\
 
                    
				\hline 
		\end{tabular}
          }
		\label{Synapse_dataset}
	\end{center}
\end{table*}

\subsubsection{Visualization} 
\begin{figure}
    \centering
    \includegraphics[scale=1]{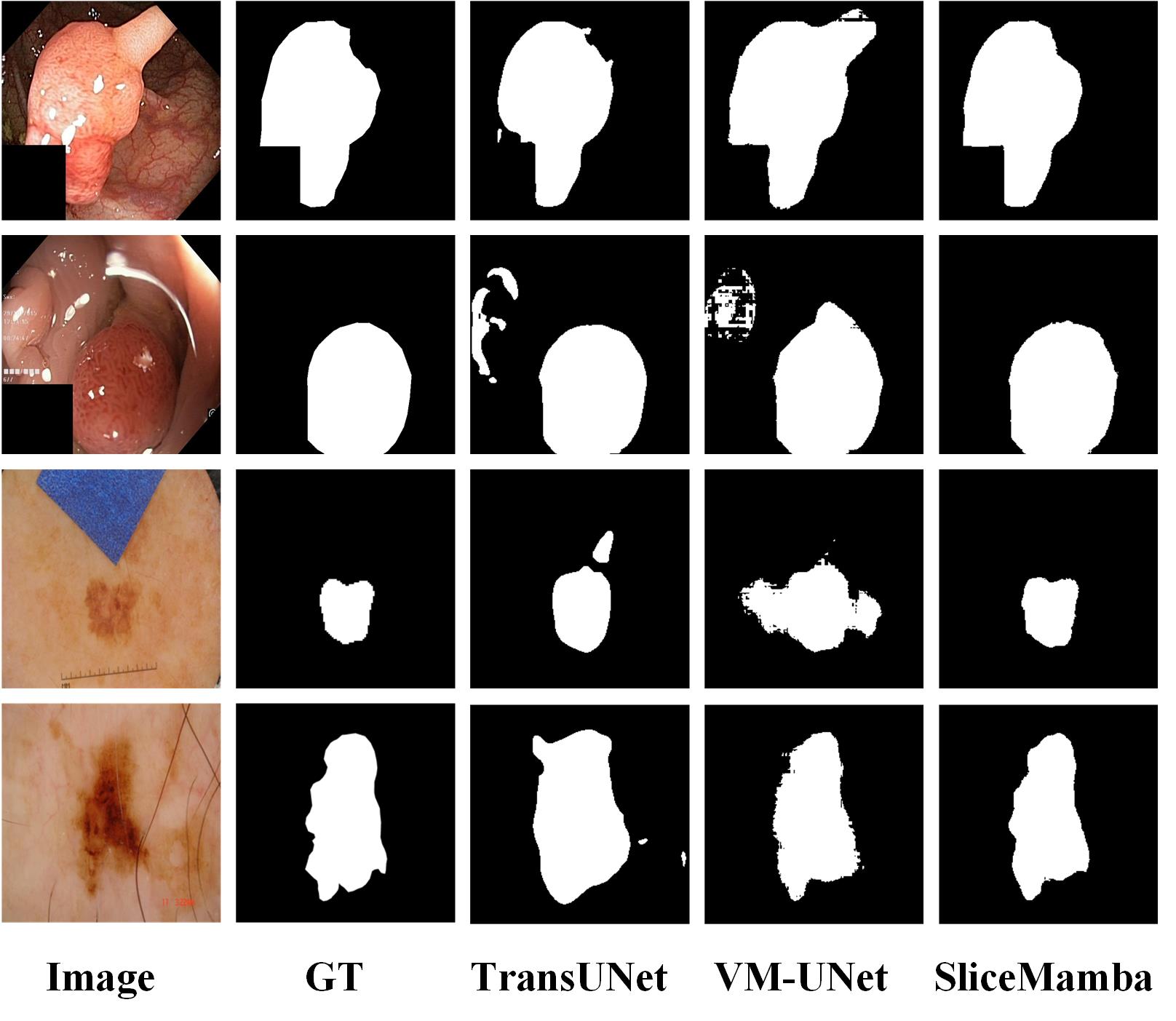}
    \caption{Visualization results on skin lesion and polyp datasets, GT stands for the ground truth segmentation mask. \textbf{Zoom in} for better details.}
    \label{fig:visual}
\end{figure}

\begin{figure}
    \centering
    \includegraphics[scale=1]{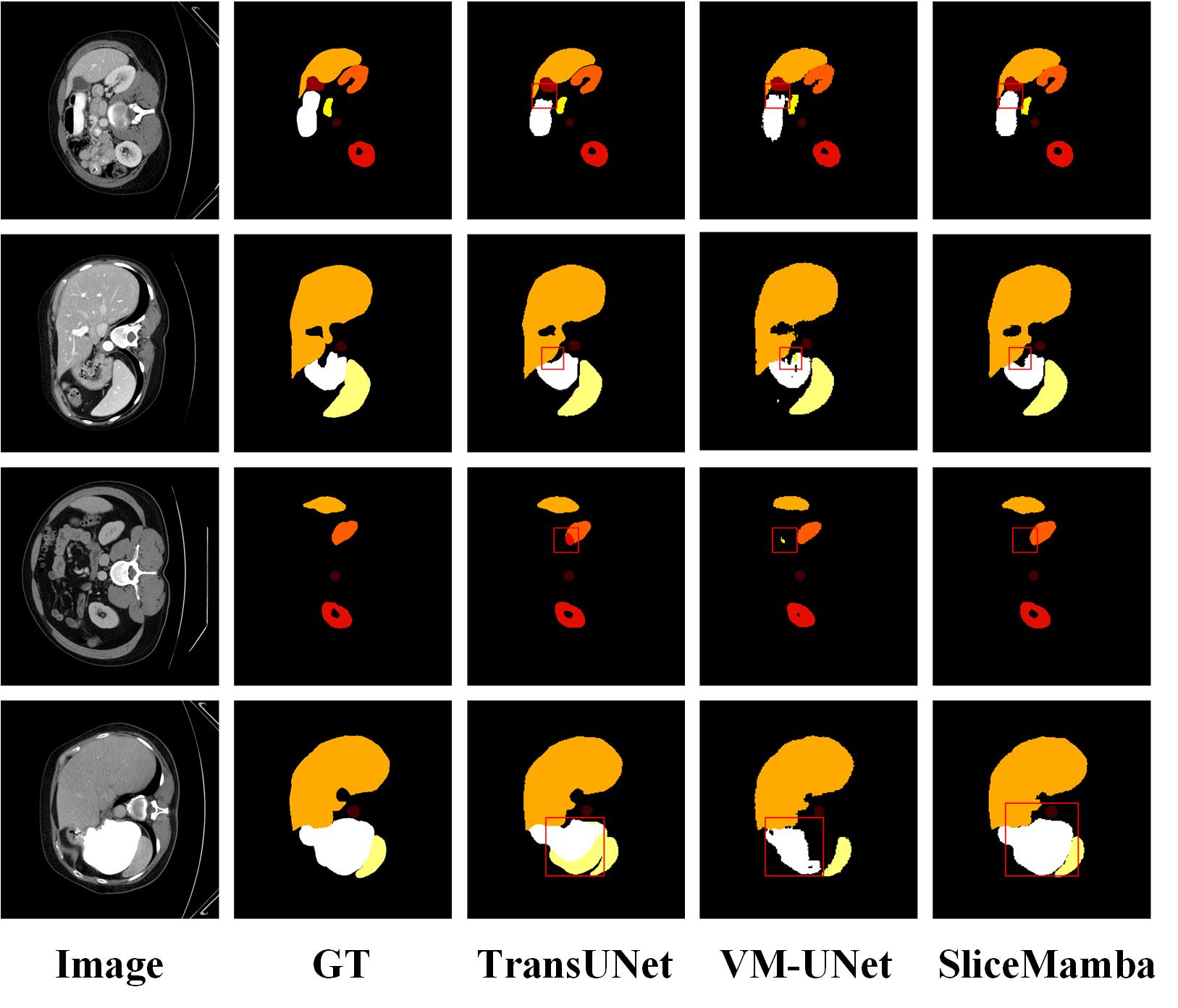}
    \caption{Visualization results on the Synapse dataset, GT stands for the Ground Truth segmentation mask. \textbf{Zoom in} for better details.}
    \label{fig:synapse_vis}
\end{figure}

As depicted in Figure.~\ref{fig:visual} and Figure.~\ref{fig:synapse_vis}, we present randomly selected visualizations of SliceMamba alongside other representative methods (the first transformer-based method, TransUNet, and the first Mamba-based method, VM-UNet) on skin lesion, polyp, and Synapse datasets.  Compared with VM-UNet, our proposed SliceMamba has better performance in segmentation details due to its strong ability to extract local features.

\subsection{Ablation Study}
In this section, we conduct a series of ablation experiments on ISIC2018 to validate the contribution of key components of SliceMamba, including the BSS module, Adaptive Slice Search (ASS), and the effect of ImageNet pre-training. We systematically analyze the impact of each component to understand how they contribute to the overall performance of the model.

\textbf{Effects of BSS module}. We take the first Mamba-based medical image segmentation model, VM-UNet (without pre-training), as the baseline and fix the feature slice method to (4,4) for the pure BSS module. As detailed in Table.~\ref{ablation}, combining the BSS module and the ASS method can bring consistent performance improvements to the baseline. To be specific, the mIoU and DSC metrics of the baseline model increase by $0.88\%$ and $0.55\%$, respectively, when equipped with the BSS module. Furthermore, when the feature slice method within the BSS module is equipped with the ASS method, its performance improves further, with mIoU and DSC metrics increasing by $1.59\%$ and $0.98\%$, respectively. From the above experiments, it can be seen that the combination of the BSS module and the ASS method can significantly enhance the performance of the baseline.

\textbf{Effects of pre-training}.
Pre-training on ImageNet and then fine-tuning on target datasets has become a key approach for deep models in natural image understanding, significantly enhancing model performance. To investigate the effect of ImageNet pre-trained parameters on medical image segmentation performance, we conducted an ablation study. As shown in Table.~\ref{pretrain}, the ISIC2018 is derived from dermoscopic imaging, which is fundamentally similar to the natural image. After loading pre-trained parameters, our method achieved improvements of $2\%$ and $1.22\%$ in the mIoU and DSC metrics, respectively. Synapse, on the other hand, is a CT imaging dataset heterogeneous to natural images. After loading pre-trained parameters, our method achieved improvements of $4.71\%$ and 6.88mm in DSC and HD95 metrics. The main reason for this significant performance boost is the small size of the Synapse dataset, where effective initialization results in more significant improvements. These findings suggest that pre-training on natural images is also beneficial for medical image tasks, particularly for small-scale medical image datasets.

\begin{table}[h]
	\begin{center}
		\caption{Ablation results of the key component of SliceMamba.}
		\scalebox{1}{
			\begin{tabular}{l rrrrc rrrrc}
				\hline 
				\multicolumn{1}{c}{\multirow{2}{*}{\centering Methods}} & \multicolumn{5}{c}{ISIC2018}  \\
                    \cmidrule(lr){2-6}
			& \multicolumn{1}{c}{mIoU$\uparrow$} & \multicolumn{1}{c}{DSC$\uparrow$} & \multicolumn{1}{c}{Acc$\uparrow$} & \multicolumn{1}{c}{Spe$\uparrow$} & \multicolumn{1}{c}{Sen$\uparrow$}  \\
				\hline 
                    baseline
                    & 78.73 & 88.10 & 94.18 & 96.03 & 88.43 \\
            
                    +BSS
                    &79.61 & 88.65 & 94.39  & 95.84 & 89.89 \\
                    
                    \rowcolor[gray]{0.9}
                    +BSS+ASS
                    &\textbf{80.32} & \textbf{89.08} & 94.73  & 96.78 & 88.33 \\
                    
				\hline 
		\end{tabular}
          }
		\label{ablation}
	\end{center}
\end{table}

\begin{table}[t]
	\begin{center}
		\caption{Ablation results with and without ImageNet pre-training.}
            \vspace{5pt}
		\scalebox{1}{
			\begin{tabular}{l cc cc}
				\hline 
				\multicolumn{1}{c}{\multirow{2}{*}{\centering Methods}} & \multicolumn{2}{c}{ISIC2018} & \multicolumn{2}{c}{Synapse} \\
                    \cmidrule(lr){2-3}\cmidrule(lr){4-5}
			& \multicolumn{1}{c}{mIoU$\uparrow$} & \multicolumn{1}{c}{DSC$\uparrow$} & \multicolumn{1}{c}{DSC$\uparrow$} & \multicolumn{1}{c}{HD95$\downarrow$}  \\
				\hline 
                    
                    ours
                    &80.32 & 89.08 & 77.24  & 22.92  \\

                    \rowcolor[gray]{0.9}
                    ours+pretrain
                    & \textbf{82.32} & \textbf{90.30} & 81.95 & 16.04 \\
				\hline 
		\end{tabular}
          }
		\label{pretrain}
	\end{center}
\end{table}

\section{CONCLUSION}
\label{sec:conclusion}
In this paper, we propose SliceMamba, a Mamba-based locally sensitive model for medical image segmentation. Specifically, we introduce the BSS module, which bi-directionally slices features and employs distinct scanning mechanisms for local feature extraction. To adapt the feature slice method for various lesions and organs, we further develop an Adaptive Slice Search method to select the optimal feature slice combination.
Based on this, we develop SliceMamba and conduct extensive experiments on various medical image segmentation datasets with diverse modalities demonstrating its efficacy. Our study highlights the significance of scanning mechanisms in Mamba-based models for medical image segmentation.

\bibliographystyle{abbrv}
\bibliography{reference}

\end{document}